\documentclass[pre,showkeys,amsmath,nofootinbib]{revtex4}

\usepackage{graphicx}
\usepackage{bm}
\usepackage{epstopdf}

\newcommand{\perc}{\%}
\newcommand{\vecphi}{\bm{\phi}}
\newcommand{\vecpi}{\bm{\pi}}

\newcommand{\vecz}{{\bf z}}

\newcommand{\lang}{\left\langle}
\newcommand{\rang}{\right\rangle}

\begin{document}

\title{Does the Boltzmann principle need a dynamical correction?}

\author{Artur B. Adib}
\email{artur_adib@brown.edu}
\affiliation{
Department of Physics, Brown University, Providence, RI 02912-1843, USA \\
Department of Physics and Astronomy, Dartmouth College, Hanover, NH 03755, USA
}

\date{\today}

\begin{abstract}
In an attempt to derive thermodynamics from classical mechanics, an approximate expression for the equilibrium temperature of a {\em finite} system has been derived [M. Bianucci, R. Mannella, B. J. West, and P. Grigolini, Phys. Rev. E {\bf 51}, 3002 (1995)] which differs from the one that follows from the Boltzmann principle $S = k \ln \Omega(E)$ via the thermodynamic relation $1/T=\partial S / \partial E$ by additional terms of ``dynamical'' character, which are argued to correct and generalize the Boltzmann principle for small systems (here $\Omega(E)$ is the area of the constant-energy surface). In the present work, the underlying definition of temperature in the Fokker-Planck formalism of Bianucci et al. is investigated and shown to coincide with an approximate form of the equipartition temperature. Its exact form, however, is strictly related to the ``volume'' entropy $S = k \ln \Phi(E)$ via the thermodynamic relation above for systems of any number of degrees of freedom ($\Phi(E)$ is the phase space volume enclosed by the constant-energy surface). This observation explains and clarifies the numerical results of Bianucci et al. and shows that a dynamical correction for either the temperature or the entropy is unnecessary, at least within the class of systems considered by those authors. Explicit analytical and numerical results for a particle coupled to a small chain ($N\sim 10$) of quartic oscillators are also provided to further illustrate these facts.
\end{abstract}


\maketitle


\section{Introduction} \label{intro}

In the notation of Planck, the connection between the thermodynamic entropy $S$ of an isolated system and its number of accessible states $W$ is often introduced as what Einstein called the {\em Boltzmann principle} \cite{cercignani98}
\begin{equation} \label{bp-w}
  S = k \ln W.
\end{equation}
For classical systems, where the number of states is replaced by continuum integrals over the phase space, two common definitions of entropy for isolated systems are\footnote{The prefactors $1/N!$ and $1/h^{3N}$ are omitted since they are irrelevant in the present context.} (see e.g. \cite{huang87}, or \cite{pearson85} and references therein)
\begin{align}
  S_{\Phi} & = k \ln \Phi(E), \label{bp-vol} \\
  S_{\Omega} & = k \ln \Omega(E), \label{bp-area}
\end{align}
where 
\begin{equation}
  \Phi(E) = \int_{H(\vecz)<E} d\vecz = \int d\vecz \, \theta[E-H(\vecz)] \\
\end{equation}
is the volume under the constant-energy surface, and
\begin{equation}
  \Omega(E) = \int_{H(\vecz) = E} d\vecz = \int d\vecz \, \delta[E-H(\vecz)]
\end{equation}
is the ``area'' of that surface. Here $H(\vecz)$ is the Hamiltonian, $\vecz$ denotes a point in the phase space, $\theta(x)$ is the Heaviside step function and $\delta(x)$ is the Dirac delta function. These two expressions for $S$ will be referred to as the volume and area entropies, respectively. It is generally acknowledged that a specific choice of $S$ above is immaterial in the thermodynamic limit \cite{huang87,pearson85}, for therein these expressions differ by terms increasingly small, but in the opposite limit of only a few degrees of freedom they lead to appreciably different observables, most notably the equilibrium temperature (given by the thermodynamic relation $1/T=\partial S/\partial E$) \cite{berd-book97}. The temperatures that follow from Eqs.~(\ref{bp-vol}) and (\ref{bp-area}) via this thermodynamic relation will be similarly referred to as the volume ($T_\Phi$) and area ($T_\Omega$) temperatures.

Almost a decade ago, Bianucci et al. (BMWG henceforth) set out to provide a ``derivation,'' and {\em generalization}, of the Boltzmann principle\footnote{It is important to stress that what these authors refer to as the Boltzmann principle is the particular case of Eq.~(\ref{bp-area}).} [Eq.~(\ref{bp-area})] with particular emphasis on systems far away from the thermodynamic limit \cite{bianucci95}. Such generalization, which will be concisely reviewed in the next section, is based upon the derivation of an approximate equation of motion for the reduced probability distribution of the degrees of freedom of interest, which is then compared with a postulated form of the Fokker-Planck equation [Eq.~(\ref{2d-fp}) below]. This comparison allows BMWG to put forward a microscopic expression for the temperature [Eq.~(\ref{general_TB}) in the present work], and additional assumptions allow them to relate their temperature expression $T_B$ with the phase space area $\Omega$ [Eq.~(\ref{bmwg-temp})]. Since the resulting relationship between $T_B$ and $\Omega$ differs from the one that follows from Eq.~(\ref{bp-area}) via the thermodynamic relation $1/T=\partial S/\partial E$ by additional terms that only vanish in the thermodynamic limit, Bianucci et al. are led to conclude that they have found a generalization of the Boltzmann principle for finite systems. This fact is confirmed by BMWG through numerical simulations of a particle coupled to a chain of oscillators, in which the authors measure the average of twice the kinetic energy of the particle and compare it with the predicted value from their proposed temperature expression.

It is clear from the above discussion that an implicit assumption in the work of BMWG is the coincidence of their analytic temperature definition  $T_B$ [Eq.~(\ref{bmwg-temp})] with $\lang v^2 \rang$ (twice the average kinetic energy, where we temporarily set the mass and the Boltzmann constant to unity), i.e. it is assumed that their temperature definition coincides with, or at least approximates the familiar equipartition one. Since it was actually the latter quantity that was used to assess the quality of the former and not the other way around, one can say that the endeavor of BMWG was to find a dynamical method for evaluating $\lang v^2 \rang$ in terms of the phase space structure of the system, and that the failure of Eq.~(\ref{bp-area}) in predicting the equilibrium temperature of their numerical experiments is related to the breakdown of the equation $\lang v^2 \rang = ( \partial S_\Omega / \partial E )^{-1}$ for finite systems. As shown in the present investigation, the BMWG temperature $T_B$ {\em is} in fact an approximation to the equipartition one, but if one assumes mixing as BMWG did in the final stages of their investigation, the asymptotic value of $\lang v^2 \rang$ becomes {\em exactly} related to $S_\Phi$ [Eq.~(\ref{vol-temp})] for any system size, establishing incidentally the equivalence between the equipartition and the volume temperatures. This shows not only that the Boltzmann principle in Eq.~(\ref{bp-area}) is indeed inappropriate to predict the equipartition temperature of small systems \cite{berd-book97}, but also that the dynamical corrections to $S_\Omega$ suggested in the work of Ref.~\cite{bianucci95} are unnecessary, since an alternate and equally legitimate form of the Boltzmann principle [Eq.~(\ref{bp-vol})] is sufficient for that purpose. This observation translates immediately into a negative answer to the question posed in the title of the present work, so long as one answers it in the context of the work of BMWG.


The remaining of this paper is organized as follows. After reviewing the formalism of BMWG (Sec.~\ref{fp-review}), it will be shown that the temperature expression found in Ref.~\cite{bianucci95} is an approximation to the usual one defined through the equipartition theorem, 
and hence that the expression derived therein as a generalization of the Boltzmann principle is but a particular approximation to the volume entropy (Sec.~\ref{fp-connection}). These findings will be illustrated by means of a Hamiltonian model similar to that of BMWG (Sec.~\ref{model}), where in addition the distribution of momenta is computed and seen to depart from the canonical one. This provides, in turn, an alternate means of explaining the inability of the area temperature to predict the average of twice the kinetic energy of a particle, as seen in the numerical experiments of Bianucci {\em et al}.

\section{The dynamical approach of BMWG} \label{fp-review}

In order to understand the origin of the dynamical generalization of the Boltzmann principle mentioned above, 
it is worth reviewing the scope and results
of BMWG. The authors of Ref.~\cite{bianucci95} were concerned with
problems whose degrees of freedom can be separated into ``system of interest'' and ``irrelevant,'' the 
latter playing the role of what is commonly known as heat bath or thermostat in the literature. However, in order 
to avoid confusion with the usual Fokker-Planck equation approach to their problem, in which the irrelevant degrees
of freedom are assumed to be in a specific macroscopic state (usually a canonical one, see e.g. \cite{ford65,zwanzig73}), the 
word {\em booster} was adopted instead. Therefore, the general goal of BMWG was to derive the thermostatistical properties of a 
system coupled to a (finite or infinite) bath whose thermal properties arise naturally from dynamics, bridging the gap 
between mechanics and thermodynamics much like the ideas of Boltzmann.

Notwithstanding the avoidance of macroscopic assumptions, two fundamental properties concerning the dynamics 
of the booster had to be assumed in order to make the problem amenable to their analytical approach, namely
(i) it is chaotic enough so that its correlation functions decay in a finite time ({\em finite correlation
time}) and (ii) the average response of 
the system to an external perturbation is linear ({\em linear response}).
In addition, the system of interest was taken to be a single particle with coordinates $x,v$ linearly coupled 
to the booster via a ``doorway'' variable $\xi$, so that $\Delta \cdot \xi$ can be interpreted as an external 
force driving the particle (or, vice versa, $\Delta \cdot x$ for the booster) whose coupling 
strength $\Delta$ has to be sufficiently small to justify their 
perturbative approach ({\em weak coupling} condition). Condition (i) implies in particular that the unperturbed
autocorrelation function of the doorway variable (with $\left<\cdot \right>_0$ indicating an equilibrium average in 
the absence of coupling), viz.
\begin{equation}
  \varphi(t) \equiv \frac{\lang \xi(t) \xi(0) \rang_0}{\lang \xi^2 \rang_0},
\end{equation}
vanishes for finite $t$ (with $\lang \xi \rang_0=0$), which in turn implies that the correlation time
\begin{equation}
  \tau \equiv \int_0^\infty \! \! dt \, \, \varphi(t)
\end{equation}
is finite. Condition (ii) states that the average behavior of the doorway variable after the introduction
of an external time-dependent perturbation $K(t)$ to its equation of motion can be written as the following
convolution (recall the assumption $\lang \xi \rang_0=0$ above)
\begin{equation}
  \lang \xi(t)\rang_K = \int_0^t du \,\, S(u) K(t-u) + \mathcal{O}(K^2),
\end{equation}
where $S(u)$ is a function that dictates how the average of $\xi$ responds to the small perturbation $K(t)$ 
(response function). Together with (i), condition (ii) furnishes two other finite quantities, the 
asymptotic susceptibility 
\begin{equation}
 \chi \equiv \lim_{t\to \infty} \chi(t) \equiv \lim_{t\to \infty} \int_0^t \! du \, \, S(u), 
\end{equation}
and the response time
\begin{equation}
  \vartheta \equiv \int_0^\infty \! \! dt \, c(t),
\end{equation}
where the convenient variable $c(t)\equiv 1 - \frac{\chi(t)}{\chi}$ was introduced so that $c(0)=1$ and $c(\infty)=0$.

Relying on the above assumptions (and also on a few others concerning the separation of time-scales) and equipped with 
a specific projection technique, Bianucci and his collaborators embarked on the rather remarkable endeavor of 
``integrating out'' the irrelevant degrees of freedom of the booster without the precise knowledge of its
Hamiltonian structure, ultimately in the search of a differential equation governing the time evolution of the
reduced distribution function of the particle, which can be formally written as \cite{bianucci95}
\begin{equation} \label{sigma}
  \sigma(x,v;t)\equiv \int d\xi d\vecpi \, \rho(x,v,\vecpi,\xi;t),
\end{equation}
where $\rho(x,v,\vecpi,\xi;t)$ is the distribution function of all the dynamical variables and $\vecpi$ 
denotes the remaining degrees of freedom of the booster. The resulting differential equation, Eq. (47) in 
Ref.~\cite{bianucci95}, is then identified with a general 2D Fokker-Planck equation of the type
\begin{eqnarray} \label{2d-fp}
  \frac{\partial}{\partial t} \sigma(x,v;t) & = & \left\{ \mathcal{L}_a^{\text{eff}} + \frac{\partial}{\partial v} A(x,v) \left( \frac{\partial}{\partial v} + \frac{mv}{kT_B} \right) \right. \nonumber \\
    & &  \left. + \frac{\partial}{\partial v} B(x,v) \left( \frac{\partial}{\partial x} + \frac{U'(x)}{kT_B} \right) \right\} \sigma(x,v;t),
\end{eqnarray}
where
\begin{equation}
  \mathcal{L}_a^{\text{eff}}=\frac{U'(x)}{m} \frac{\partial}{\partial v} - v \frac{\partial}{\partial x}
\end{equation}
is the effective time evolution operator (Liouvillian) of the unperturbed particle. From a comparison between
the coefficients of their equation and the structure of the above equation, they suggest (despite the implicit dependence
on the particle coordinates $x,v$) that the following expression should be identified with the temperature of
the booster (Eq. (52) in Ref.~\cite{bianucci95}):
\begin{equation} \label{general_TB}
  kT_B(x,v) \equiv \frac{\lang \xi^2 \rang_0}{\chi} \frac{\int_0^\infty \! du \, \varphi(u) \left[ \frac{\partial}{\partial x} x_a(t-u) \right] }
                           {\int_0^\infty \! du \, c(u) \left[ \frac{\partial}{\partial x} x_a(t-u) \right] },
\end{equation}
where $x_a(t-u)$ is the unperturbed evolution of the position of the particle ``backwards in time'' (being essentially 
the responsible for the implicit dependence of $T_B$ on $x,v$). The subscript in $T_B$ stands for BMWG. To render the above equation independent 
of $x,v$, BMWG consider the following three cases: (a) the particle is a harmonic oscillator with $V(x)=m\omega^2 x^2/2$, 
(b) $\varphi(u)=c(u)$, and (c) a ``natural time scale'' (NTS) condition exists separating the 
typical timescale of the system of interest from the relaxation TS of the booster (the former being much greater
than the latter). Under condition (a), the temperature reduces to
\begin{equation}
  kT_B = \frac{\lang \xi^2 \rang_0}{\chi} \frac{\text{Re} \left[ \hat\varphi(\omega) \right] }{\text{Re} \left[ \hat c(\omega) \right]},
\end{equation}
with hats indicating a Fourier transform, whereas in the case of NTS (c) one obtains
\begin{equation} \label{bmwg-temp-NTS}
  kT_B = \frac{\lang \xi^2 \rang_0 \tau}{\chi \vartheta},
\end{equation}
which coincides with the one obtained within a Langevin approach (cf. Eq. (22) in Ref.~\cite{bianucci95}), as 
expected from the dichotomy of the time scales. Condition (b), however, is peculiar in the sense that it 
recovers the usual canonical results, which (with the exception
of making the connection with Kubo's linear response theory clear) was not particularly relevant for the remaining 
part of their work.

To draw a parallel with Boltzmann's principle, BMWG invoked {\em mixing} for the booster, in which case the time-dependent
susceptibility can be found via ``geometrical'' arguments to assume the form \cite{bianucci94}
\begin{equation} \label{chi}
  \chi(t) = \frac{1}{\Omega_b(E)} \frac{\partial}{\partial E} \left\{ \Omega_b(E) \lang \xi^2 \rang_0 [1- \varphi(t)] \right\},
\end{equation}
where $\Omega_b(E)$ is the phase space area of the unperturbed booster with energy $E$. With this result in hand, one can
write, for the case (c) above for example,
\begin{equation} \label{bmwg-temp}
  \frac{1}{kT_B} = \frac{\partial}{\partial E} \ln \Omega_b(E) + \frac{\partial}{\partial E} \ln(\lang \xi^2 \rang_0 \tau),
\end{equation}
whereas for case (a) one simply replaces $\tau$ with $\text{Re} \left[ \hat\varphi(\omega) \right]$ in this expression. 
This result should be contrasted with the one obtained from what BMWG calls the Boltzmann principle [i.e. Eq.~(\ref{bp-area})], namely
\begin{equation}  \label{omega-temp}
  \frac{1}{kT_\Omega} = \frac{\partial}{\partial E} \ln \Omega(E).
\end{equation}
The additional term in Eq.~(\ref{bmwg-temp}), called by BMWG a ``dynamical correction'' to Eq.~(\ref{omega-temp}), is then argued to become negligible for sufficiently large systems, 
thus recovering the result above and causing no conflict with the standard macroscopic thermostatistics of Boltzmann, i.e.
the Boltzmann principle has been ``generalized'' to cover systems with only a few degrees of freedom. In the following section it is
shown that this generalization is actually an approximate form of the volume temperature $T_\Phi$.

\section{Connection with the volume temperature and entropy} \label{fp-connection}

Besides heuristic arguments of units, it is not clear how one can identify the ``constant'' $T_B$ [Eq. (\ref{general_TB})] 
obtained by BMWG through a comparison with Eq. (\ref{2d-fp}) as an actual temperature. In the usual derivation of the Fokker-Planck 
equation (or, equivalently, of the Langevin equation) for a Brownian particle \cite{reif65,reichl98}, the 
temperature comes into play by invoking the equipartition theorem, viz. $kT\equiv m\langle v^2 \rangle$
(this is indeed the case for the Langevin treatment of BMWG, cf. Eq. (22) in Ref.~\cite{bianucci95}), but in the
case of the Fokker-Planck equation derived by BMWG or the generic one in Eq. (\ref{2d-fp}), 
both with $x,v$-dependent coefficients, it is difficult to see whether their temperature definition is compatible 
with equipartition. Nevertheless, after the simplifications introduced by any of the cases (a), (b), or (c) of the previous
section (the only ones considered by BMWG in their generalization of the Boltzmann principle),
one can show that {\em the definition of $kT_B$ coincides with an approximate form of $m\langle v^2 \rangle$}. Indeed, focusing on 
condition (c) for notational simplicity (the remaining cases (a) and (b) follow by a trivial modification of the procedure below), the 
time-independent differential equation satisfied by $\tilde \sigma(x,v)$ obtained via the projection method reduces to (cf. Eq. (47) of 
Ref.~\cite{bianucci95})
\begin{align}
  0 = & -v\frac{\partial \tilde\sigma}{\partial x} + 
    \frac{V'(x)}{m} \frac{\partial \tilde\sigma}{\partial v} - 
    \frac{\Delta^2 x \chi}{m} \frac{\partial \tilde\sigma}{\partial v} +
    \frac{\Delta^2 \lang \xi^2 \rang_0 \tau}{m^2} \frac{\partial^2 \tilde\sigma}{\partial v^2} \nonumber \\
    & + \frac{\Delta^2 \lang \xi^2 \rang_0 \eta^2}{m^2} \frac{\partial^2 \tilde\sigma}{\partial v \partial x} +
    \frac{\Delta^2 \chi \vartheta}{m} \frac{\partial}{\partial v}(v\tilde\sigma) \nonumber \\
    & + \frac{\Delta^2 \chi \beta^2}{m} \frac{\partial \tilde\sigma}{\partial v},
\end{align}
where the constants $\eta^2$ and $\beta^2$ are defined in Eqs. (6) and (13) of Ref.~\cite{bianucci95}.
The tilde indicates that the distribution here is {\em not} identical to Eq. (\ref{sigma}) due to the 
approximations carried out in the actual process of integration, or ``projection'' (see below).
Multiplying the above equation through by $v^2$ and integrating over $x,v$, most terms disappear under the 
assumption that $\tilde \sigma(x,v)\to 0$ sufficiently fast for $x,v\to \infty$ and 
$\langle x \rangle_{\tilde\sigma}=\langle v \rangle_{\tilde\sigma}=0$ (for simplicity I assume the average 
position of the particle is centered at the origin, though it is not difficult to adapt 
this derivation for the general case), so one is left with
\begin{equation} \label{mv2}
  0 = \frac{2 \Delta^2 \lang \xi^2 \rang_0 \tau}{m^2} - \frac{2 \Delta^2 \chi \vartheta}{m} \langle v^2 \rangle_{\tilde \sigma},
\end{equation}
where $\langle \cdot \rangle_{\tilde \sigma} \equiv \int_{-\infty}^{\infty}dx \int_{-\infty}^{\infty}dv \, \cdot \, \tilde \sigma(x,v)$.
Solving this equation for $m\lang v^2 \rang_{\tilde \sigma}$ we obtain the right hand side of Eq. (\ref{bmwg-temp-NTS}), 
i.e. $m\lang v^2 \rang_{\tilde \sigma}=kT_B$. This is the desired connection between $T_B$ and the equipartition temperature, {\em q.e.d.} 

It is important to emphasize, however, that the reduced density $\tilde{\sigma}$ (and hence the aforementioned connection) is only approximate. This can be seen in the discussion of paragraph 1, page 3005 of Ref.~\cite{bianucci95} -- apart from conditions (i) and (ii) of the previous section, it is always assumed that there is a large separation of relaxation and intrinsic time scales of the booster. This condition, which amounts to an assumption of weak coupling, was used explicitly between Eqs.~(39) and (40) of Ref.~\cite{bianucci95} (note also that Eq.~(39) of Ref.~\cite{bianucci95} is itself a perturbative result). We now note that the {\em exact} average $m \langle v^2 \rangle_\sigma$, 
obtained with $\sigma$ [Eq. (\ref{sigma})] before the approximations carried out in the projection method of BMWG which lead to 
$\tilde{\sigma}$, furnishes the following expression
\begin{equation} \label{exact-v2}
  m \langle v^2 \rangle_\sigma = m \int \! \! dx dv d\xi d\vecpi \, v^2 \, \frac{\delta[E-H(x,v,\xi,\vecpi)]}{\Omega(E)},
\end{equation}
where $H(x,v,\xi,\vecpi)$ is the Hamiltonian of the system + booster, and
\begin{equation}
  \rho(x,v,\xi,\vecpi)=\frac{\delta[E-H(x,v,\xi,\vecpi)]}{\Omega(E)}
\end{equation}
is the asymptotic solution of the Liouville equation for mixing systems (microcanonical distribution, cf. \cite{bianucci94}), in 
accordance with the assumption of mixing in Sec. IV-D of Ref.~\cite{bianucci95}. By means of simple integral manipulations used 
in the generalized equipartition theorem (see e.g. \cite{huang87,berd-book97}), we can rewrite Eq. (\ref{exact-v2}) in terms
of the volume entropy $S_\Phi=k\ln\Phi(E)$ as
\begin{equation} \label{vol-temp}
  m \lang v^2 \rang_\sigma = \left[ \frac{\partial}{\partial E} \ln \Phi(E) \right]^{-1} = k \left( \frac{\partial S_\Phi}{\partial E} \right)^{-1} \equiv kT_\Phi.
\end{equation}
Observe that, contrary to the approximate conclusion made after Eq.~(\ref{mv2}), the above equation establishes an {\em exact} equivalence between the equipartition and the volume temperatures, as already anticipated in Sec.~\ref{intro}, showing that the knowledge of the phase space volume alone (or $S_\Phi$) is sufficient to predict the equilibrium value of $m \lang v^2 \rang_\sigma$. Though not new \cite{huang87,berd-book97}, this result renders the dynamical corrections introduced in Ref.~\cite{bianucci95} unnecessary. This fact is illustrated in the next section by means of explicit analytical and numerical calculations.

\section{Quartic oscillators: An analytically tractable model} \label{model}

In this last section a specific Hamiltonian model will be analyzed both analytically and numerically. Since BMWG relied on a similar 
model to illustrate their findings, the present analysis serves the similar purpose of illustrating on an equal footing the results 
established in the last section. It also shows that, for such small systems where Eqs.~(\ref{bp-vol}) and (\ref{bp-area}) yield appreciably different results, a noncanonical distribution of momenta is a necessary and fundamental property, in contrast to what is argued in Ref.~\cite{bianucci95}.
 
The model adopted herein is described by the Hamiltonian structure
\begin{equation} \label{hamilt-bmwg}
  H(\vecz) = H_1(x,p) + H_2(\vecphi, \vecpi) + g \, H_{1,2}(x,\phi_0),
\end{equation}
where $x,p$ and $\vecphi=(\phi_0,\ldots,\phi_{N-1})$, $\vecpi=(\pi_0,\ldots,\pi_{N-1})$
are the particle and booster canonically conjugate coordinates, respectively, $g$ is a
controllable coupling constant and
\begin{align}
  H_1(x,p) & = \frac{p^2}{2} + \frac{1}{2} \, \omega^2  x^2 \label{hamilt-x}, \\
  H_{2}(\vecphi,\vecpi) & = \sum_{i=0}^{N-1} \frac{\pi^2_i}{2} + \sum_{i=1}^{N-1} \left[ \frac{\phi_i^4}{4} + \frac{(\phi_i-\phi_{i-1})^4}{4} \right], \label{hamilt-adib} \\
  H_{1,2}(x,\phi_0) & = x \, \phi_0, \label{hamilt-coup}
\end{align}
are the particle, booster and coupling Hamiltonians, respectively.
The intrinsic time scale of the particle is set by $\omega^{-1}$, whereas its coupling with the booster
is set linearly through the ``doorway'' variable $\phi_0$. 
It should be noted that this model can be rendered unphysical
when $\omega \rightarrow 0$ (i.e. in the free particle limit) since the energy would no longer
be bounded from below (due to the remaining linear coupling term $H_{1,2}$) and thence no stable thermodynamical state would exist
\cite{ford88} (that shall not concern us here since $\omega$ will always be non-zero).

The essential difference between the above
model and the one adopted by BMWG is the absence of harmonic terms in the booster Hamiltonian, which permits a simple 
analytic treatment based on scaling arguments, as we shall see in the next sections. While not particularly relevant for the present analysis, 
it will also be shown that this model retains most of the quantitative features of the model of Ref.~\cite{bianucci95}
(see, in particular, $q$ and $\mathcal{E}$ below). More important than these quantitative similarities
is its mixing behavior which, just as in the model of BMWG, is suggested by the numerical experiments through 
the observed finite relaxation time towards a unique equilibrium state. 

For the numerical simulations to be 
reported soon, the equations of motion were evolved with an optimized
fourth-order symplectic integrator \cite{mclachlan92}, resulting in numerical energy fluctuations
$\Delta E/E$ smaller than one part in $10^6$. Initial conditions were picked randomly from the
density $\rho(x,p,\vecphi,\vecpi;t=0)=\delta(x) \delta(p) \delta(H_2 - E)$ \cite{bianucci95}, and no dependence
of the observables in equilibrium on these configurations was observed (being already a good indication
of ergodicity by Birkhoff's theorem \cite{berd-book97}). The typical integration
time used was $t=10^8$, after allowing a suitable relaxation time of order $t_{rel}=10^7$. Since we are not
interested in nonequilibrium features, no ensemble average was done (the coincidence of time and phase space
averages relying on the Birkhoff-Khinchin theorem for ergodic systems \cite{berd-book97}). For simplicity,
the harmonic frequency of the particle, $\omega$, was set to unity (in contrast to the approach of BMWG,
the specific value of $\omega$ is immaterial for the validity of the method adopted here).

\subsection{Momentum distribution} \label{dist}

As a first probe of ergodicity, one can test whether the observed histogram of the
momentum of the particle $f(p)$ approximates the distribution law obtained from the structure function
$\Omega_{p}(E)$ of the remaining degrees of freedom $x$, $\vecphi$ and $\vecpi$ (see below), i.e. for
large enough integration times one expects
\begin{equation} \label{phase-dist}
  f(p) \to \frac{\Omega_{p}(E-p^2/2)}{\Omega(E)},
\end{equation}
where $\Omega(E)$ is the structure function of the whole system.
Most quantities of interest here, including $\Omega_{p}$, can be obtained from the ($2N+1$)-dimensional
phase space volume
\begin{equation} \label{gammax2}
  \Phi_{p}(E) = \int \! d x \, d \vecphi \, d \vecpi \, \, \theta\!\left(E - [x^2/2 + H_2(\vecphi,\vecpi)]\right),
\end{equation}
where $\theta$ is the unit step function and the coupling Hamiltonian $H_{1,2}$ was neglected
({\em weak coupling approximation}). In fact, assuming the above volume to be bounded and sufficiently smooth for 
all $E>0$ (such assumption will be adopted throughout this section), the phase space area $\Omega_{p}$ can be obtained 
from $\Phi_{p}$ via the general relation \cite{khinchin49} 
\begin{equation} \label{omega-gamma}
  \Omega_i(E)=\frac{d \Phi_i(E)}{d E},
\end{equation}
where $i$ labels the desired (sub)volume. Though evaluating Eq. (\ref{gammax2}) can be a hard
task for general Hamiltonians, for the choice presented in Eqs. (\ref{hamilt-x})-(\ref{hamilt-coup}) 
a simple scaling argument can be used to show that (cf. Ref.~\cite{adib02a}):
\begin{equation} \label{gamma2-scaling}
  \Phi_{p}(E) = \tilde{a} \, E^{(3N+2)/4},
\end{equation}
where henceforth the symbols $\tilde{a},\tilde{b},$ etc. will refer to any function that does not
depend on $E$. Equation (\ref{omega-gamma}) then gives
\begin{equation} \label{omega2-scaling}
  \Omega_{p}(E) = \tilde{b} \, E^{(3N-2)/4},
\end{equation}
and therefore, by inserting this relation in Eq. (\ref{phase-dist}), one gets the desired distribution law:
\begin{equation} \label{final-distlaw}
  f(p) = \frac{1}{Z} \left( 1 - \frac{1}{E} \frac{p^2}{2} \right)^{(3N-2)/4},
\end{equation}
where $Z$ is a normalization constant. This expression
is valid for any $N>0$, provided the dynamics of the system is sufficiently ergodic
(recall that $f(p)$ was defined as a histogram). Note also that in the
limit $N\to \infty$ with $E/N$ constant one recovers the usual Boltzmann factor $\exp(-\beta p^2/2)$,
with $\beta=3N/4E$.

\begin{figure}
\includegraphics[width=3in]{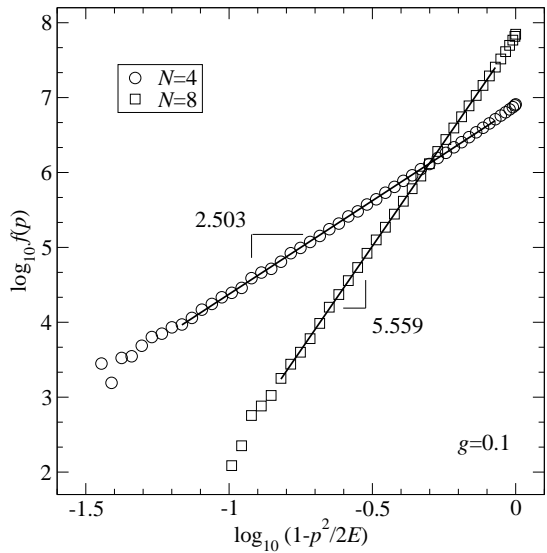}
\caption{ \label{f_vs_p}   Logarithmic plot of the histograms $f(p)$ obtained from the numerical simulations for $N=4,8$ and
  $g=0.1$. The slope of each curve is indicated and should be compared with the values
  predicted by Eq. (\ref{final-distlaw}), viz. $2.5$ ($N=4$) and $5.5$ ($N=8$). The finite-time sampling manifests
  here through the departure from the predicted power-law at the tail of the
  distributions, which nevertheless tends to smooth out as $t\to \infty$.
}
\end{figure}
Figure \ref{f_vs_p} shows the momentum distribution function of the harmonic oscillator obtained from
the numerical simulations for the booster model in Eq. (\ref{hamilt-adib}), from which one obtains good 
evidence for ergodicity given its clear agreement with Eq. (\ref{final-distlaw}). The 
``departure from Gaussianity'' introduced by BMWG was also computed, namely
$q = 1-\langle p^4\rangle/[3\langle p^2\rangle^2]$, to show not only that the distribution is
close to canonical (see below however), but also that its numerical value is very close to the one obtained from
the model of Ref.~\cite{bianucci95}. Indeed, the observed value of $q$ in the present model
was $q=0.1255$ for $N=8$, which should be compared to $q\approx 0.1$ obtained by BMWG for the same number of
particles.

In the following section we will be observing the temperature of the system via the average momentum square
of the particle. Here I show that the mismatch between $T_\Phi$ and $T_\Omega$ to be found in that section
can alternatively be attributed to the noncanonicity of $f(p)$. Indeed, using the integral
\begin{equation}
  \int_{-1/\sqrt{a}}^{1/\sqrt{a}} p^n \left( 1 - a p^2 \right)^b dp= 
    \frac{\left( 1 + (-1)^n \right)}{2 a^{\frac{1}{2}(n+1)}} \cdot
     \frac{\Gamma(b+1) \Gamma\left(\frac{1+n}{2}\right) } { \Gamma\left( \frac{3}{2} + b + \frac{n}{2} \right) },
     \quad (a,b>0)
\end{equation}
where $\Gamma(u)$ is the gamma function and $n=0,1,2,\ldots$, we obtain, after using the property 
$\Gamma(u+1)=u\,\Gamma(u)$ and reading the values of $a,b$ from Eq. (\ref{final-distlaw}),
\begin{equation}
  \lang p^2 \rang = \int_{-\sqrt{2E}}^{\sqrt{2E}} p^2 f(p) \, dp= \frac{4E}{3N+4}
\end{equation}
(compare with Eq. (\ref{t_phi}) below). Note that this result differs from the one obtained using the 
Boltzmann factor, namely $\lang p^2 \rang = 4E/3N$ (compare Eq. (\ref{t_omega})). This difference, though 
small for large $N$, is ultimately the motivation for
BMWG to conclude that their corrected temperature is indeed an improvement over the ``standard'' one ($T_\Omega$,
obtained from Boltzmann's principle), which shows additionally
that, contrary to the discussion that accompanies Fig. 6 of Ref.~\cite{bianucci95}, the small departure from 
canonicity of the momentum distribution is a pertinent property of these small systems.

\subsection{Temperature} \label{BP}

Using the same weak coupling condition adopted in the last section, one can neglect the interaction Hamiltonian
and compute both the phase space volume and area (this time for the whole system) through a procedure
similar to the one leading to Eqs. (\ref{gamma2-scaling}) and (\ref{omega2-scaling}). One then finds,
respectively,
\begin{align}
  \Phi(E) & = \tilde{c} \, E^{(3N+4)/4}, \label{gamma-scaling} \\
  \Omega(E) & = \tilde{d} \, E^{3N/4}, \label{omega-scaling}
\end{align}
from which the temperatures $T_\Phi$ and $T_\Omega$ follow trivially, viz.
\begin{align}
  T_\Phi & = \left(\frac{\partial \ln\Phi(E)}{\partial E}\right)^{-1} = \frac{4E}{3N + 4} = \frac{4 \epsilon}{3} \, \frac{1}{1+4/3N}, \label{t_phi} \\
  T_\Omega & = \left(\frac{\partial \ln\Omega(E)}{\partial E}\right)^{-1} = \frac{4E}{3N} = \frac{4\epsilon}{3}, \label{t_omega}
\end{align}
where $\epsilon \equiv E/N$ is the total energy per degree of freedom and $k$ was set to unity (this harmless
simplification will be adopted throughout the remaining part of the paper). 
In what follows, the {\em numerically observed temperature} $T_{obs}$ is defined as the time-average
$\overline{v^2}$, which should coincide with the ensemble average $\lang v^2 \rang$ [and hence with $T_\Phi$, cf. Eq.~(\ref{vol-temp})] for ergodic systems.

\begin{figure}
\includegraphics[width=3in]{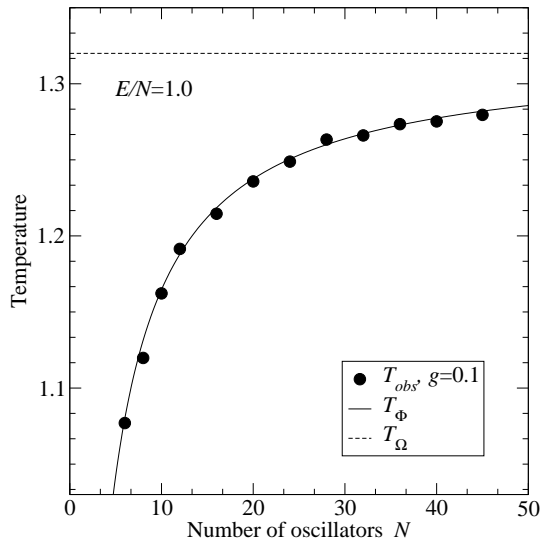}
\caption{ \label{T_vs_N} Comparison between the observed temperatures of the particle $T_{obs}$ from the
  numerical simulations (full circles) and from the analytic predictions, Eqs. (\ref{t_phi})
  and (\ref{t_omega}) (continuous and dashed lines, respectively).
  The observed equilibrium temperature $T_{obs}$ was obtained from the asymptotic value
  of the time-average $\overline{v^2}$ after allowing a suitable relaxation
  and observation time (see text).
}
\end{figure}
In Fig. \ref{T_vs_N} the observed temperatures from the numerical simulations with $\epsilon=1.0$,
$g=0.1$ and various values of $N$ are presented, as well as the analytic predictions Eqs. (\ref{t_phi})
and (\ref{t_omega}). The agreement with $T_\Phi$ is evident, and we can see clearly
from this figure that both temperatures should coincide in the thermodynamical limit \cite{berd-book97}.
In order to verify that these results can in fact explain the discrepancy found by BMWG between the ``standard''
form of Boltzmann's principle [which gives rise to Eq. (\ref{omega-temp})] and their numerical simulations, the relative error
${\mathcal E} = (T_{obs}-T_\Omega) /T_\Omega$ was computed both for the data
available here and in that reference (the temperatures from the latter are $T_\Omega \approx 14.95$ and 
$T_{obs} \approx 11.78$), giving ${\mathcal E}_A \approx -15.2\,\perc$ and 
${\mathcal E}_B \approx -21.2\,\perc$,
respectively. Note that not only the magnitude of these relative deviations are quite close, but also their
{\em sign} is the same, i.e. the general trend of $T_\Omega$ is to {\em overestimate} the final
equilibrium temperature, just as BMWG found (see also \cite{bannur97}). The additional error
${\mathcal E}_A-{\mathcal E}_B \approx 6\, \perc$ can be further explained from the fact that
the phase space area in the approximate temperature expression of BMWG involves the degrees of freedom 
of the booster only, neglecting the contribution of $x$ and $p$ which tends to increase $T_\Omega$ 
and thus to decrease ${\mathcal E}$. This is the source of the ``$10\perc$'' deviation
from the observed temperature mentioned before the conclusions of Ref.~\cite{bianucci95}.

\section{Conclusions}

It is important to emphasize that the purpose of the present contribution was {\em not} to determine whether the volume or the area entropy is the correct starting point of thermostatistics for small systems. Although the former bears an important thermodynamic property that is independent of the system size (known as ``adiabatic invariance,'' cf. \cite{adib02b} and references therein), these very small {\em and} isolated systems such as the ones studied by BMWG are usually far removed from experience, and we do not know whether our empirical and macroscopic thermodynamics acquires a different form in this case. Experiments with small clusters of atoms is a field that can potentially settle this question \cite{freeman96}, and a theoretical comparison between the two entropies above has already been carried out in this context \cite{jellinek00}. 

Quite apart from elucidating the ad hoc definition of the BMWG temperature [cf. discussion following Eq.~(\ref{mv2})], the present contribution has established that one does not have to resort to dynamical quantities such as time correlation functions in order to predict the equilibrium temperature of a small system. Such idea was put forward in the work of Ref.~\cite{bianucci95} in the form of a correction to the temperature that follows from Eq.~(\ref{bp-area}) for the particular case of a system composed of a particle coupled to a set of ``irrelevant'' degrees of freedom, leading its authors to conclude that a correction for the Boltzmann principle is necessary when one is dealing with small systems. The key identity that escaped the attention of BMWG is Eq.~(\ref{vol-temp}), which establishes an {\em exact} relation between the equipartition temperature and the phase space volume, a quantity that can be computed much in the same way as the usual phase space area. This shows that, as long as we are working with the assumptions of Ref.~\cite{bianucci95}, the particular form of the Boltzmann principle in Eq.~(\ref{bp-vol}) is sufficient for predicting the equipartition temperature, without any dynamical correction.

\acknowledgments
The author is indebted to Prof. Oliver Penrose and the anonymous referee for their constructive and detailed comments, which largely reshaped the original manuscript. Correspondences or private discussions with Professors Victor Berdichevsky, David Montgomery, Murilo Almeida and Paolo Grigolini are also acknowledged. Financial support and computational resources (the latter through the Field Theory/Cosmology group) were partially provided by the Dartmouth College.

\end{document}